\documentclass[journal,letterpaper]{IEEEtran}
\usepackage{amsmath,amssymb,amsthm}
\usepackage{microtype}
\usepackage{enumitem}

\usepackage{mathtools}

\usepackage{nth}

\usepackage{booktabs}
\usepackage{graphicx,xcolor}
\usepackage{hyperref}

\newtheorem{theorem}{Theorem}

\theoremstyle{definition}
\newtheorem{definition}{Definition}

\usepackage[utf8]{inputenc}    
\usepackage[T1]{fontenc}       

\theoremstyle{remark}

\newcommand{\bR}{\ensuremath\mathbb{R}}
\newcommand{\bx}{\ensuremath\boldsymbol x}
\newcommand{\bu}{\ensuremath\boldsymbol u}
\newcommand{\by}{\ensuremath\boldsymbol y}
\newcommand{\cX}{\ensuremath\mathcal{X}}
\newcommand{\cU}{\ensuremath\mathcal{U}}
\newcommand{\cY}{\ensuremath\mathcal{Y}}
\newcommand{\cL}{\ensuremath\mathcal{L}}

\long\def\Note#1{\bgroup\color{red}#1\egroup}

\makeatletter
\providecommand*{\diff}%
	{\@ifnextchar^{\DIfF}{\DIfF^{}}}
\def\DIfF^#1{%
	\mathop{\mathrm{\mathstrut d}}%
		\nolimits^{#1}\gobblespace}
\def\gobblespace{%
		\futurelet\diffarg\opspace}
\def\opspace{%
	\let\DiffSpace\!%
	\ifx\diffarg(%
		\let\DiffSpace\relax
	\else
		\ifx\diffarg[%
			\let\DiffSpace\relax
		\else
			\ifx\diffarg\{%
				\let\DiffSpace\relax
		\fi\fi\fi\DiffSpace}
\makeatother

\usepackage{tikz}
\usetikzlibrary{shapes,arrows}
\tikzstyle{block} = [draw, fill=blue!20, rectangle, 
    minimum height=2.5em, minimum width=5em]
\tikzstyle{sum} = [draw, fill=blue!20, circle, node distance=.8cm]
\tikzstyle{input} = [coordinate]
\tikzstyle{output} = [coordinate]
\tikzstyle{pinstyle} = [pin edge={to-,thin,black}]
\usetikzlibrary{positioning}
\usetikzlibrary{patterns}
\usetikzlibrary{calc}
\usetikzlibrary{shapes.misc}
\usetikzlibrary{backgrounds}

\usepackage{pgfplots}
\pgfplotsset{compat=newest}
\pgfplotsset{plot coordinates/math parser=false}
\tikzstyle{block} = [draw, fill=white, rectangle, 
    minimum height=1.6em, minimum width=3.2em]
\tikzstyle{sum} = [draw, circle, node distance=1cm]
\usepackage[style=ieee,backend=biber,doi=false,isbn=false]{biblatex}
\usepackage{balance}\balance

\title{Recent Advances in Analysis and Design of Cyber-physical Systems using Passivity Indices}
\author{Hasan Zakeri%
\thanks{\quad}
\thanks{Department of Electrical Engineering, University of Notre Dame, USA. \url{hzakeri@nd.edu}, \url{antsaklis.1@nd.edu}} and Panos J. Antsaklis
\thanks{The partial support of ARO under Grant No. ARL~W911NF-17-1-0072  is gratefully acknowledged.}}

\begin{document}

\maketitle
\begin{abstract}
Analysis and resilient design of Cyber-physical Systems have greatly benefited from energy based concepts of passivity and dissipativity. Recently, there has been much research devoted to the use of passivity indices in different components of Cyber-physical systems. Passivity indices are measures of passivity, indicating how passive a system is or how far is it from being passive and generalize passivity based methods to systems that might not be passive. In this paper, we will review recent advances in the use of passivity indices in Cyber-physical systems. We will overview how the indices have been defined and applied to different components of Cyber-physical systems and how they are used in the resilient design of compositional Cyber-physical systems. 
\end{abstract}

\section{Introduction}
A crucial component of Cyber-physical Systems (CPS) is the tight integration between the  Cyber and the Physical parts. CPS usually consists of many components working together, each with different models and requirements. ``Although the diversity of models and formalisms supports a component-based “divide and conquer” approach to CPS development, it poses a serious problem for verifying the overall correctness and safety of designs at the system level.'' Compositionality of the design is one of the major challenges in Cyber-physical Systems (CPS). 
CPS design requires reliability and robustness both at the same time
\autocite{lee_cyber_2008}.

Passivity, and more generally dissipativity, use a general notion of energy to derive abstractions of dynamical systems regardless. Such abstractions provide a common framework to model different components of the system and they have shown great promise in the analysis and resilient design of CPS as recently reviewed in~\autocite{passivity.cyber,sztipanovits_toward_2012}. 

Passivity indices, as measures of passivity, express how passive a system is or how far a system is from being passive. The indices generalize the notion of passivity to systems that are not passive. Naturally, methods based on passivity indices generalize to a broader class of systems. Passivity indices framework has attracted much attention in CPS research and has been applied to many problems beyond the scope of passivity and dissipativity. 
In this paper, we will review some of the recent advances in analysis and design of CPS using passivity indices, and particularly, we summarize how the indices are defined for different components in CPS and how they can be used in the analysis and resilient design of such systems. In the next section, we review preliminaries and basic definitions of passivity indices. \autoref{sec:design} covers a design method to passivate a system. Next, in \autoref{sec:NCS}, we will discuss how passivity indices are defined and used in networked control systems. \autoref{sec:hybrid} will cover systems involving both continuous and discrete dynamics and how the indices are defined and used in switched and hybrid systems. \autoref{sec:conc} concludes the paper.

\section{Preliminaries}\label{sec:pre}
Consider a  continuous-time dynamical system \(\mathbf H:\bu\to\by\), where \(\bu\in\cU\subseteq\bR^m\) denotes the input and \(\by\in\cY\subseteq\bR^p\) denotes the corresponding output. There exists a real-valued function \(w(\bu(t),\by(t))\) (often written as \(w(t)\) when clear from content) associated with \(\mathbf H\), such that for all input and output pairs of \(\bu(t)\) and \(\by(t)\) of the system, and for any \(t_1\geq t_0,\) 
\begin{equation}
	\int\limits_{t_0}^{t_1}\vert w(t)\vert\diff t<\infty
\end{equation}
for any \(t_1\geq t_0.\)
The function \(w\) is called a \emph{supply rate function.} 
Consider a continuous-time system described by
\begin{equation}\label{eq:system}
	\begin{aligned}
		\dot \bx&=f(\bx,\bu)\\
		\by&=h(\bx,\bu),
	\end{aligned}
\end{equation}
where \(f(\cdot,\cdot)\) and \(h(\cdot,\cdot)\) are Lipschitz mappings of proper dimensions, and assume the origin is an equilibrium point of the system; i.e., \(f(0,0)=0\) and \(h(0,0)=0.\) 

\begin{definition}\label{def:dissipativity}
The system described by~\eqref{eq:system} is called \emph{dissipative with respect to supply rate function \(w(\bu(t),\by(t)),\)} if there exists a nonnegative real-valued scalar function \(V(\bx),\) called the \emph{storage function,} such that \(V(0)=0\) and for all \(\bx_0\in\cX,\) all \(t_1\geq t_0,\) and all \(\bu\in\bR^m\) 
\begin{equation}\label{eq:dissipativity}
	V(\bx(t_1))\leq V(\bx(t_0))+\int\limits_{t_0}^{t_1}w(\bu(t),\by(t))\diff t.
\end{equation}
where \(\bx(t_0)=x_0\) and \(\bx(t_1)\) is the state at \(t_1\) resulting from initial condition \(x_0\) and input function \(u(\cdot).\)
\emph{The dissipation inequality~\eqref{eq:dissipativity}} expresses the fact that the energy ``stored'' in the system at any time \(t\) is not more than the initially stored energy plus the total energy supplied to the system by its input during this time. 
\end{definition}
If \(V(\bx)\) is differentiable, then this is equivalent to 
\begin{equation}\label{eq:dissipativitydiff}
	\dot V(\bx)\triangleq\frac{\partial V}{\partial\bx}\cdot f(\bx,\bu)\leq w(\bu(t),\by(t)).
\end{equation}
\begin{definition}\label{def:passivity}
The system~\eqref{eq:system} is called \emph{passive,} if it is dissipative with respect to the supply rate function \(w(\bu,\by)=\bu^\intercal\by.\)
\end{definition}
Passivity can be applied to linear and nonlinear systems with a generalized notion of an energy storage function. Passivity is particularly appealing in CPS design because it is preserved when systems are combined in parallel or feedback~\autocite{process.passivity}. Passivity imposes additional restrictions on a system than stability. For example, linear SISO passive systems aren't simply Lyapunov stable but are also minimum phase and have a low relative degree. 

Passivity indices are introduced as measures of passivity and they extend passivity based tools to non-passive systems as well.
\begin{definition}[Input Feed-forward Passivity Index]\label{def:IFP_index}
The system~\eqref{eq:system} is called \emph{Input Feed-forward Passive (IFP)} if it is dissipative with respect to supply rate function \(w(\bu,\by)=\bu^\intercal\by-\nu\bu^\intercal\bu\) for some \(\nu\in\bR,\) denoted as IFP(\(\nu\)). Input feed-forward passivity index for system~\eqref{eq:system} is the largest \(\nu\) for which the system is IFP. 
\end{definition}
IFP index is equivalent to the largest gain that can be put in a negative feed-forward interconnection with the system such that the overall system is passive.
\begin{definition}[Output Feedback Passivity]\label{def:OFP_index}
The system~\eqref{eq:system} is called \emph{Output Feedback Passive (OFP)} if it is dissipative with respect to supply rate function \(w(\bu,\by)=\bu^\intercal\by-\rho\by^\intercal\by\) for some \(\rho\in\bR,\) denoted as OFP(\(\rho\)). Output feedback passivity index for system~\eqref{eq:system} is the largest \(\rho\) for which the system is OFP. 
\end{definition}
OFP index is the largest gain that can be placed in positive feedback with a system such that the interconnected system is passive. If either one of the indices for a system is positive, we say that the system has an ``excess of passivity,'' and similarly, if either one is negative, we say the system has a ``shortage of passivity.''

When applying the two indices simultaneously, a system is said to have OFP(\(\rho\)) and IFP(\(\nu\)), or IF-OFP(\(\rho,\nu\)), based on the following dissipation inequality:
\begin{equation*}
    \int\limits_0^T[(1+\rho\nu)\bu^\intercal\by-\rho\by^\intercal\by-\nu\bu^\intercal\bu]\diff t\geq V(\bx(T))-V(\bx(0)).
\end{equation*}
When \(\rho=0\) and \(\nu=0\) the passivity index condition reduces to the definition of passivity.
\begin{theorem}[\autocite{Khalil,hill.moylan.76}]
Output feedback passive systems with \(\rho>0\) are \(\cL_2\) stable. Moreover, if system \(G\) is strictly passive with OFP index \(\rho,\) then \(G\) is finite-gain \(\cL_2\) stable with gain \(\gamma\leq\frac1\rho.\)
\end{theorem}
\begin{theorem}[\autocite{mccourt_dissipativity_2013}]\label{thm:stability:indices:LMI}
Consider the interconnection of two nonlinear systems (2.1). Assume that the two systems in the interconnection have indices \((\rho_i,\nu_i).\) The interconnection is \(\cL_2\) stable if the following matrix is positive definite:
\begin{equation}
    A=\begin{bmatrix}(\rho_1+\nu_2)I&\frac12(\rho_1\nu_1-\rho_2\nu_2)I\\
        \frac12(\rho_1\nu_1-\rho_2\nu_2)I&(\rho_2+\nu_1)I
    \end{bmatrix}>0
\end{equation}
\end{theorem}
For a review of relationship between passivity, passivity indices, conic systems, and positive real systems, see~\autocite{mccourt_connection_2009,meng.spr}.
Passivity indices under operationa limitations for nonlinear systems as well as approximate methods to find them are presented in~\autocite{zakeri_passivity_2019}. Local passivity indices for nonlinear systems and sum of squares methods to find the local indices are also presented in~\autocite{zakeri.acc2016}.

\section{Indices in Passivation and Design}\label{sec:design}
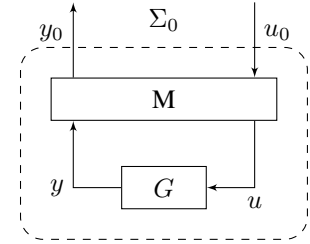
\begin{figure}[hbt]
    \centering
    \begin{tikzpicture}[auto, node distance=1cm,>=latex']
        \node [block, minimum width = 3cm, fill=white] (M) {M};
        \node [block, below = .6cm of M, fill=white] (G) {\(G\)};
        \draw [->] (G.west) -| node {$y$}  ($(M.south west)!0.1!(M.south east)$);
        \draw [->] ($(M.south west)!0.9!(M.south east)$) |- node {$u$} (G.east);
        \draw [->] ($($(M.north west)!0.9!(M.north east)$)+(0,1)$) -- 
            node[pos=.4] {$u_0$} ($(M.north west)!0.9!(M.north east)$);
        \draw [->] ($(M.north west)!0.1!(M.north east)$) --
            node[pos=.6] {$y_0$} ($($(M.north west)!0.1!(M.north east)$)+(0,1)$);
        \coordinate (A) at ($(M.south east)+(0.4,0)$);
        \coordinate (B) at ($(G.south east)+(0,-0.4)$);
        \begin{scope}[on background layer]
            \draw[rounded corners=3mm, dashed, label=Rec] ($(M.north west)+(-0.4,0.4)$) rectangle (B -| A);
            \node [above  = 0.5cm of M] {$\Sigma_0$};
        \end{scope}
    \end{tikzpicture}
    \caption{M-matrix interconnection}
    \label{fig:Mmatrix}
\end{figure}

Passivity indices can be manipulated by series, feedback, or parallel interconnection. A generalization of these methods is given in~\autocite{xia_passivity_2014} by using an input-output transformation matrix. Appropriate design of this matrix, called \emph{the M-matrix,} guarantees positive passivity levels for the system. This transformation matrix allows the use of a non-passive controller to guarantee the passivity and stability of a feedback controller. Consider the system \(G\) and a general input-output transformation matrix \(M\) as shown in \autoref{fig:Mmatrix}.
The matrix \(M\) is considered to be invertible and defined as 
\begin{equation}
    M\triangleq\begin{bmatrix}
        m_{11}I&m_{12}I\\m_{21}I&m_{22}I.
    \end{bmatrix}
\end{equation}
It is shown in~\cite{xia_passivity_2014} that the passivity indices of the system \(\Sigma_0:\bu_0\to\by_0\) depend on the gain \(\gamma\) of system \(G\) and the elements of \(M,\) as stated in \autoref{thm:Mmatrix}.
\begin{theorem}\label{thm:Mmatrix}
    Consider a finite gain stable system \(G\) with gain \(\gamma\) and a passivation matrix \(M\) as shown in \autoref{fig:Mmatrix}. The system \(\Sigma_0:\bu_0\to\by_0\) is 
    \begin{enumerate}
        \item passive, if \(M\) is chosen such that
        \begin{equation}
            m_{11}=m_{21},\quad m_{22}=-m_{21},\quad m_{11}\geq m_{22}\gamma>0.
        \end{equation}
        \item OFP with OFP level \(\rho_0=\frac12\left(\frac{m_{11}}{m_{21}}+\frac{m_{12}}{m_{22}}\right)>0,\) if 
        \begin{equation}
            m_{21}\geq m_{22}\gamma>0,\quad m_{11}m_{22}>m_{12}m_{21}>0.
        \end{equation}
        \item IFP with IFP level \(\nu_0=\frac12\left(\frac{m_{21}}{m_{11}}+\frac{m_{22}}{m_{12}}\right)>0,\) if
        \begin{equation}
            m_{11}\geq m_{12}\gamma>0,\quad m_{12}m_{21}>m_{11}m_{22}>0.
        \end{equation}
        \item IF-OFP with passivity indices \(\delta_0=\frac12\frac{m_{11}}{m_{21}}>0\) and \(\epsilon_0=\frac a2\frac{m_{21}}{m_{11}}>0,\) if
        \begin{equation}
            m_{11}>0,\quad m_{12}=0,\quad m_{21}\geq\frac{m_{22}\gamma}{\sqrt{1-a}}>0,
        \end{equation}
        where \(0<a<1\) is an arbitrary real number.
    \end{enumerate}
\end{theorem}
Conditions are given in \autocite{xia_passivity_2014} for a feedback interconnection of passivated system \(\Sigma_0\) and another arbitrary system. For proof of \autoref{thm:Mmatrix}, see \autocite{xia_passivity_2014-1}, where the same passivation method is applied to a human controller. A quasi-linear model of human controller is presented in \autocite{mcruer_human_1959}. It is well-known that passivity and stability of a closed-loop system may not be preserved in the presence of communication effects, as shown in the next section. 

The M-matrix method is applied as a human/machine interface design in~\autocite{xia_guaranteeing_2015}. Human controllers are modeled as linear time delay systems, which are known to be non-passive. This method is applied to a human in the loop to guarantee positive passivity indices and passivate the closed-loop system.

Passivity indices are used in~\autocite{zakeri_data-driven_2019-1} to derive a new data-adriven fault identification and controller reconfiguration algorithm. The algorithm relies only on the system's input and output data, and does not require a detailed system description. This algorithm can be readily applied to various applications, including Cyber-attack mitigation, without significant modifications or tuning.

\section{Passivity Indices in Networked Control Systems}\label{sec:NCS}
Networks are an essential part of CPS, making possible the integration between many distributed components. The elements are usually spatially distributed and communicate over a wired or wireless link. The use of communication links in large scale systems brings many advantages including ease of maintenance, flexibility, scalability, and lower costs. However, employing shared multi-purpose communication networks in the core of Cyber-physical systems, as opposed to traditionally dedicated connections, brings out many challenges, making Networked Control Systems research of significant importance~\autocite{ge_distributed_2017,baillieul_control_2007,baillieul_guest_2003,antsaklis_networked_2004,zhang_network-induced_2013}.
Packet dropouts~\autocite{su_robust_2018}, signal quantization~\autocite{gu_networked_2014}, time delays~\autocite{liu_networked_2014,heemels_networked_2010}, fading channels~\autocite{su_design_2018}, and limited information transfer~\autocite{zhang_resilient_2016} are among many challenges addressed in the literature~\autocite{grubmuller_influence_2018}.


Delays are the most common effect of communication networks, and it is known that delay can disrupt passivity~\autocite{niemeyer_stable_1991}. In~\autocite{kottenstette_stable_2007}, the authors employed a wave variable transformation to compensate for the effects of delays when interconnecting passive systems. A modified wave variable transformation is applied to the case of networked switched systems in~\autocite{mccourt_stability_2010}, where two switched systems connected in negative feedback communicate over a network with time delays. The non-passive nature of the delay can render the interconnection non-passive. This framework is extended in~\autocite{agarwal_dissipativity_2016-1} to the network interconnection of hybrid systems. The main idea in this transformation is to treat the delayed network as a 2-port network. If this two-port network is passive, then it ensures the passivity of the whole interconnection. The interconnection, including the transformations, is depicted in \autoref{fig:wvt}. The wave variable transformation is given as
\begin{equation}
    \begin{aligned}
        \begin{bmatrix}u_1\\\hat v_1 \end{bmatrix}=\frac1{\sqrt{2b}}
        \begin{bmatrix}I&bI\\-I&bI\end{bmatrix}\begin{bmatrix}y_d^{(2)}\\y^{(1)} \end{bmatrix}\\
        \begin{bmatrix}\hat u_2\\\hat v_2\end{bmatrix}=\frac1{\sqrt{2b}}
        \begin{bmatrix}I&bI\\-I&bI\end{bmatrix}\begin{bmatrix}y^{(2)}\\y_d^{(1)} \end{bmatrix}
    \end{aligned}
\end{equation}
where
\begin{equation}
    \hat u_2=u_1(t-T_2)=u_2,\ \hat v_1=v_2(t-T_1)=v_1
\end{equation}
where the network delays \(T_1\) and \(T_2\) and the design parameter \(b\) are constant. For time varying delays, the following modified transformation is proposed
\begin{equation}
    \begin{aligned}
        \hat u_2&=g_1(t)u_1(t-T_2(t))=g_1(t)u_2(t)\\
        \hat v_1&=g_2(t)v_2(t-T_1(t))=g_2(t)v_1(t)
    \end{aligned}
\end{equation}
where the design functions \(g_1(t),g_2(t)\) and delays \(T_1(t),T_2(t)\) satisfy
\begin{gather*}
    g_1^2(t)\leq1-\frac{\diff T_2}{\diff t},\ g_2^2(t)\leq1-\frac{\diff T_1}{\diff t}\\
    \frac{\diff T_1}{\diff t}\leq 1,\ \frac{\diff T_2}{\diff t}\leq 1
\end{gather*}
It is shown in~\autocite{agarwal_dissipativity_2016,agarwal_dissipativity_2016-1} that the interconnection of two passive hybrid automata connected over a network with time-varying or constant delay using the modified wave variable transform is passive, and under certain assumptions, stable. Details on the passivity of hybrid systems can be found in \autoref{sec:hybrid}.

Passivity of a networked control system in the presence of packet dropouts is studied in~\autocite{wang_feedback_2015}. In this framework, the system is modeled as a discrete-time switched nonlinear system that switches between two modes---an uncontrolled mode in which the system evolves open loop, and a controlled mode in which a control input is applied to the system---is analyzed. If the ratio of the time steps for which the system evolves open loop versus the time steps for which the system evolves closed loop is bounded below a critical ratio, then the nonlinear system is locally passive in this sense.

Quantization is another widely seen phenomenon in digital controllers and communication channels. The proposed control framework of~\autocite{zhu_passivity_2012} maintains passivity for switched and non-switched systems under quantization, based on the use of an input-output coordinate transformation to recover the passivity property.
\begin{figure}
    \centering
        \begin{tikzpicture}[auto, node distance=.75cm,>=latex']
	      \node [block, minimum width = 4cm] (M) {WVT};
	      \node [block, above of=M, node distance=1.2cm] (system) {$\Sigma_1$};
	      \node [block, below  = 1.8cm of M, minimum width = 4cm] (M2) {WVT};
	      \node [block, below of=M2, node distance=1.2cm] (feedback) {\(\Sigma_2\)};
	      \coordinate (A) at ($(M.north west)!0.1!(M.north east)$);
	      \coordinate (B) at ($(M.north west)!0.9!(M.north east)$);
	      \draw (A |- system) node [sum] (S1) {};
	      \node [block, below = 1.2cm of A] (D1) {\(T_1(t)\)};
	      \node [block, below = 1.2cm of B] (D2) {\(T_2(t)\)};
	      \draw (B |- feedback) node [sum] (S2) {};
	      \node [input, left = 1.2cm of S1] (input1) {};
	      \node [input, right= 1.2cm of S2] (input2) {};
	      \draw (input1 |- feedback.west) node [input] (out2) {};
	      \draw (input2 |- system.east) node [input] (out1) {};
	      \draw [->] (input1) -- node {$w_1$} node [pos=.9] {+} (S1);
	      \draw [->] (input2) -- node {$w_2$} node [pos=.9] {+} (S2);
	      \draw [->] (A) -- node {$y_{2d}$} node [pos=.9] {-} (S1);
	      \draw [->] (M2.south -| B) -| node[pos=.66] {$y_{1d}$} node [pos=.9] {+} (S2);
	      \draw [->] (S1) -- node {$e_1$} (system);
	      \draw [->] (S2) -- node {$e_2$} (feedback);
	      \draw [->] (feedback) -- node[pos=.66] {$y_2$} (out2);
	      \draw [->] (system) -- node[pos=.66] {$y_1$} (out1);
	      \draw [->] (feedback.west -| A) -- (A |- M2.south); 
	      \draw [->] (B |- out1) -- (B);
	      \draw [->] (D1) -- node {$v_1$} (A |- M.south);
	      \draw [->] (B |- M.south) -- node {$u_1$} (D2);
	      \draw [->] (D2) -- node {$u_2$} (M2.north -| B);
	      \draw [->] (M2.north -| A) -- node {$v_2$} (D1);
    \end{tikzpicture}
    \caption{Two systems connected over a network with delays using wave variable transform}
    \label{fig:wvt}
\end{figure}

\subsection{Event-triggered Networked Control Systems}
Event-triggered control is an efficient network control method that limits the communication loads on the network and provides a balance between performance, communication constraints, and control actions~\autocite{lemmon_event-triggered_2010}. 
In a typical event-triggered feedback framework, the information, whether the measurement or the control action, is transmitted only when necessary. Compared to time-driven control where constant sampling period is applied to guarantee stability in the worst case scenario, the possibility of reducing the number of computations and thus of transmissions while guaranteeing desired levels of performance makes event-triggered control very appealing in networked control CPS. Passivity and passivity indices based control combined with event-triggered networked control systems (NCS) provide a powerful platform for designing CPS. This section is dedicated to recent advances in passivity based event-triggered networked control design.

The stability conditions for an event-triggered networked system based on the passivity properties of sub-systems have been explored in~\autocite{yu_event-triggered_2011,zhu_passivity_2017,zhu_passivity_2014-1}, with focus on passivity and passivation of event-triggered feedback interconnected systems of two input-feedforward, output-feedback passive systems. It is shown that passivity indices of continuous feedback systems can be determined from the passivity levels of individual subsystems. The passivation conditions to render a non-passive plant passive are also obtained based on passivity indices. The results can be viewed as the extension of the well-known compositional property of passivity to event-triggered systems. The proposed event-triggering condition guarantees that these indices can be achieved. The passivation conditions depend on the passivity indices of the plant and controller and the event-triggering condition, which shows the trade-off between desired performance and communication resource utilization.

In~\autocite{rahnama_qsr-dissipativity_2016,rahnama_qsr-dissipativity_2018}, QSR-dissipativity, passivity and finite-gain \(\cL2\)-stability conditions for an event-triggered NCS are proposed, where an input-output event-triggering sampler condition is located on the plant's output side, the controller's output side, or both sides, leading to a considerable decrease in communication load between sub-units in NCS. Passivity and stability conditions depend on passivity levels for the plant and controller. The results illustrate the trade-off among passivity levels, stable performance, and systems dependence on the rate of communication between the plant and controller. 

Many of the challenges mentioned in the previous section for network control systems, without event triggered framework, can persist into event-triggered scheme as well. 
The formation control problem of networked passive systems with event-triggered communication is studied in~\autocite{yu_distributed_2012}, where a triggering condition to achieve distance-based formation among the agents with an ideal network model is derived. In~\autocite{yu_quantized_2012}, the quantized output synchronization problem of networked passive systems with event-driven communication is studied to achieve an event-driven communication strategy such that output synchronization errors of the networked passive systems are bounded by the quantization errors of the signals transmitted in the communication network.

The triggering condition is derived in~\autocite{yu_event-triggered_2013} in the presence of network-induced delays. The results are based on the passivity theorem which allows characterization of a large class of output feedback stabilizing controllers and can consider network induced delays both from the plant to the network controller and from the network controller to the plant. Additionally, the effects of quantization of the transmitted signals in the communication network are taken into consideration and conditions for finite-gain \(\cL_2\)-stability are achieved in the presence of time-varying (or constant) network induced delays with bounded jitters, without requiring that the network induced delays are upper bounded by the inter-event time.

\begin{figure}
    \centering
    \includegraphics[width=\linewidth]{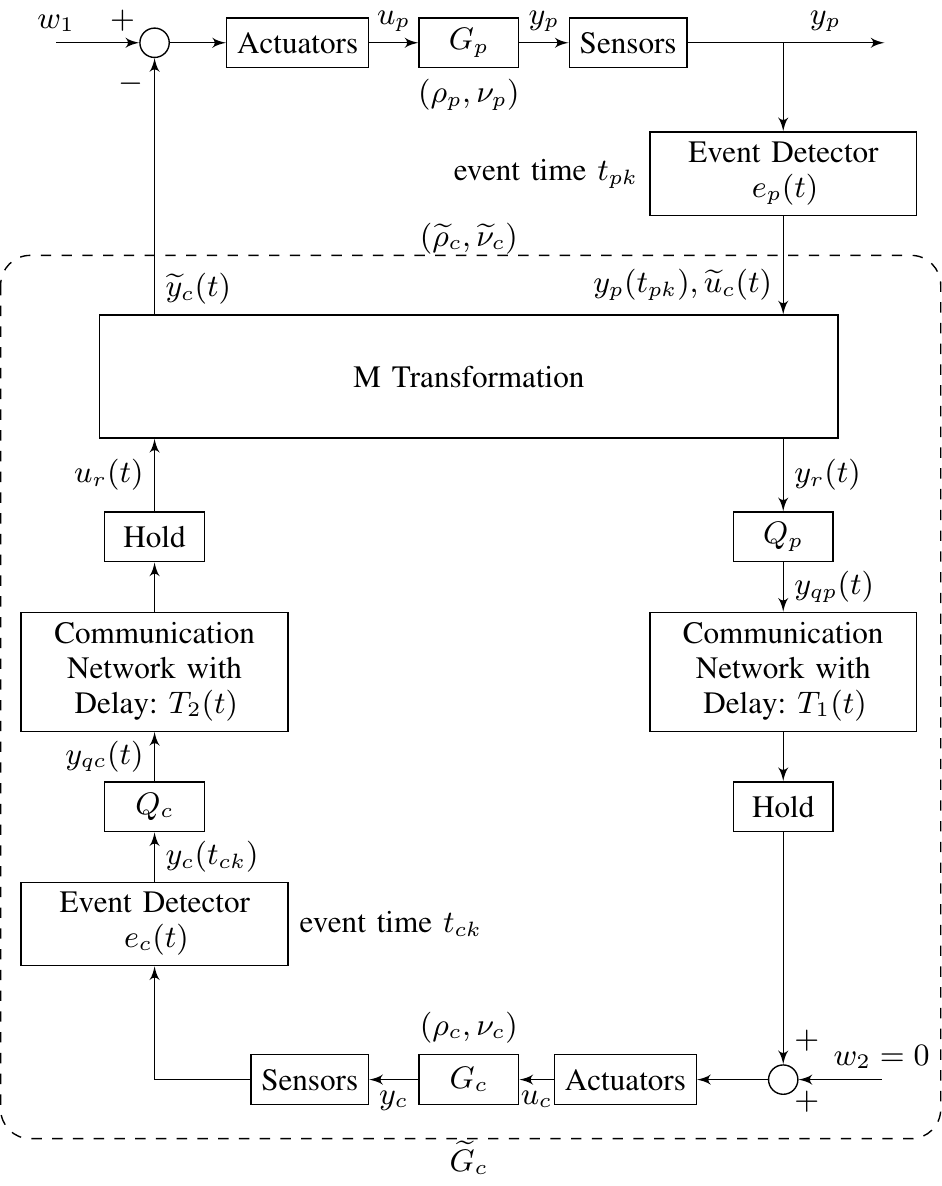}
    \caption{ A Networked Control System Interconnection of two IF-OFP systems with Quantization, Time-Varying Delays and Event-Detectors on both sides}
    \label{fig:event}
\end{figure}

In \autocite{rahnama_passivity-based_2018}, design framework that considers the effects of network induced time-delays, signal quantization and data loss in communication links between the plant and controller, as illustrated in \autoref{fig:event}, is introduced, along with conditions for \(\cL2\)-stability and robustness for the control design. The proposed asynchronous triggering conditions do not rely on the exact knowledge of the systems dynamics and are located on both sides of the communication network. A lower-bounds on inter-event time intervals for the triggering conditions and characterize the design's robustness against external disturbances is also proposed. For a detailed illustration of the relationship between stability, robustness and passivity levels of plant and controller, as well as an analysis of robustness against packet dropouts and passivity levels for the entire event-triggered networked control system, see~\autocite{rahnama_passivity-based_2018}.
\section{Passivity Indices in Switched and Hybrid Systems}\label{sec:hybrid}
Unlike continuous systems, a switched system has an unusual phenomenon that must be taken into consideration when dealing with change of energy~\autocite{zhao_dissipativity_2008}. Analyzing the stability of switched systems is not a trivial task, and with the complexity of the dynamics involved.

The applicability of passivity has been extended when defined for switched systems. Definitions of passivity for switched systems have been proposed in~\autocite{zefran_notion_2001,zhao_notion_2006,zhao_dissipativity_2008}.
These definitions of passivity assume that under arbitrary switching, each subsystem is passive. This is an important assumption that guarantees passivity for the trivial case of no switching.  Aside from that, different definitions impose different assumptions on the system regarding the energy added due to switching.

Passivity-based control of hybrid systems with a common storage function has attracted many applications in electrical systems. For instance, \autocite{escobar_hybrid_1998} applied the hybrid passivity idea to a three-phase voltage-sourced reversible-boost-type rectifier, and~\autocite{espinosa-perez_passivity-based_2004} proposed a passivity-based control strategy for switched reluctance motors with nonlinear magnetic circuits. A switching control strategy based on energy is presented in~\autocite{zhao_hybrid_2001} to globally stabilize the cart-pendulum system. 
However, the definition of passivity through common storage function, while useful, might not cover all practical consideration of a system. Each subsystem might have its own physical meaning, and its own storage function accordingly.
Passivity, based on a common storage function is similar to the classic definition os passivity and is fully discussed in~\autocite{haddad_dissipativity_2001,haddad_non-linear_2001}. Here, we will focus on passivity and dissipativity of switched systems with multiple storage functions (sometimes referred to as \emph{decomposable dissipativity}).

Consider the switched system of the form
\begin{equation}\label{eq:switched_system}
    \begin{aligned}
        \dot\bx&=f_\sigma(\bx,\bu_\sigma),\\
        \by&=h_\sigma(\bx)
    \end{aligned}
\end{equation}
where \(\sigma\) is the switching signal taking values in 
\[
    M=\{1,2,\dots,s\}
\]
which may depend on time or state, or both, or even be generated by higher level hybrid feedback in the loop. The switching signal \(\sigma\) can be characterized by the switching sequence
\begin{equation}
    \Sigma=\{\bx_0;(i_0,t_0),(i_1,t_1),\dots,(i_r,t_r),\dots\mid i_r\in M,n\in\mathbb N\}
\end{equation}
where \(t_0\) is the initial time, \(\bx_0\) is the initial state, and \(\mathbb N\) is the set of nonnegative integers. When \(t\in[t_k,t_{k+1}), \sigma(t) = i_k ,\) that is, the \(i_k\)th subsystem is active. We assume that the state of the switched system~\eqref{eq:switched_system} does not jump at the switching instants, i.e., the trajectory \(\bx(t)\) is everywhere continuous. 

The first work to extend the passivity theory to switched systems through multiple storage functions is~\autocite{zefran_notion_2001} requires each storage function to be nonincreasing on the switching sequence of consecutive ``switched on'' times as a prerequisite to meet the nonincreasing condition of multiple Lyapunov functions, which, in turn, guarantees stability~\autocite{branicky_multiple_1998}. It also requires that each subsystem is passive ``on average'' while inactive.

The following present another definition of passivity with a relaxed set of assumptions which still leads to stability. 
\begin{definition}[{\autocite{zhao_notion_2006}}]
A switched system is passive with a set of storage functions \(V_i(\bx)\) and functions \(\alpha_{ijt},\gamma_j\) such that the following statements are true
\begin{enumerate}
    \item  The energy a system accumulates while inactive is bounded,
    \begin{equation}
        V_i(\bx(t_2))\leq\alpha_{ji(t_2-t_1}(V_i(\bx(t_1)))+\int\limits_{t_1}^{t_2}\bu^\intercal\by\diff t.
    \end{equation}
    \item The composition of these functions \(\alpha_{jit}\) is bounded,
    \begin{equation}
        \alpha_{ji_1\tau_1}\circ\alpha_{ji_2\tau_2}\circ\dots\circ\alpha_{ji_k\tau_k}(t)\leq\gamma_j(t).
    \end{equation}
\end{enumerate}
\end{definition}
The first condition of the definition implies that when each subsystem is active, then it is passive. It is explicitly shown that when the system stays in a single subsystem, the first condition reduces to the traditional definition of passivity. However, \autocite{zhao_notion_2006}~does not discuss passivity of the negative or parallel feedback interconnection. 
This definition counts energy accumulated while a subsystem is inactive at a different, typically reduced, rate than when the subsystem is active. This amounts to each inactive system having a reduced ``imported energy'' from the active system. With this adjustment, each subsystem can tolerate a larger energy supply \(\bu^\intercal\by\) and still maintain passivity. Two sets of functions are introduced in~\autocite{zhao_notion_2006} to account for this reduced storage rate, however, the significance of these functions as well as how to compute them is not straightforward.

In~\autocite{zhao_dissipativity_2008}, passivity for switched systems is defined and used to show stability results. Specifically, it was shown that passive systems are Lyapunov stable, that negative feedback induces asymptotic stability, and that output strictly passive systems are asymptotically stable.
\begin{definition}[\autocite{zhao_dissipativity_2008}]\label{def:dissipativity_switched}
System~\eqref{eq:switched_system} is said to be dissipative under the switching law \(\Sigma\) if there exist positive-definite continuous functions \(V_1(\bx),V_2(\bx),\dots,V_s(\bx),\) called storage functions, locally integrable functions \(w_i^i(\bu_i,h_i),1\leq i\leq s,\) called supply rates, and locally integrable functions \(w_j^i(\bx,\bu_i,h_i,t),1\leq i,j\leq s,i\neq j,\) called cross-supply rates, such that the following conditions hold
\begin{enumerate}
    \item Each subsystem \(i\) is dissipative with respect to \(w_i^i\) while active, i.e.,
    \begin{equation*}
        V_{i_k}(\bx(t))-V_{i_k}(\bx(r))\leq\int\limits_{t_1}^{t_2}w_{i_k}^{i_k}(\bu(\tau),h_{i_k}(\bx(\tau))\diff\tau
    \end{equation*}
    \item Each subsystem \(j\) is dissipative with respect to \(w_j^i\) when the \(i\)th subsystem is active, i.e., for \(k=0,1,2,\dots,t_k\leq r\leq t<t_{k+1},\)
    \begin{multline}
        V_j(\bx(t))-V_j(\bx(r))\leq\\\int\limits_r^tw_j^{i_k}(\bx(\tau),u_{i_k}(\tau),h_{i_k}(\bx(\tau)),\tau)\diff\tau,
    \end{multline}
    \item for \(j\neq i_k,k=0,1,2,\dots,t_k\leq r\leq t<t_{k+1},\) and for any \(i,j,\) there exist \(\bu_i(t)=\alpha_i(\bx(t),t)\) and \(\varphi_j^i(t)\in\cL_1^+[0,\infty),\) which may depend on \(\bu_i\) and the switching sequence \(\Sigma,\) such that 
\begin{align}
    f_i(0,\alpha_i(0,t))&\leq0,&&\forall t\geq t_0\\
    w_i^i(\bu_i(t),h_i(\bx(t)))&\leq0&&\forall t\geq t_0
\end{align}
and 
\begin{equation*}
    w_j^i(\bx(t),\bu_i(t),h_i(\bx(t)),t)-\phi_j^i(t)\leq0,\ \forall j\neq i,\forall t\leq t_0.    
\end{equation*}
\end{enumerate}
\end{definition}
In this framework, because all subsystems share the same state variables, the storage function of subsystems are still changing on the time intervals when they are inactive. The active subsystem drives the state which, in turn, causes the change of the storage functions of inactive subsystems. This ``changing'' energy of any inactive subsystem, though not necessarily real energy, is viewed as ``exported energy'' from the active subsystem, and is characterized by cross-supply rates. Unlike the classical definition of dissipativity, in~\autocite{zhao_dissipativity_2008}, positive definite storage functions are required to induce stability and output stabilization. This requirement is the same as that of multiple Lyapunov functions~\autocite{branicky_multiple_1998}.
Passivity comes from this definition by setting
\begin{equation*}
    w_j^j(\bu_j,h_j)=\bu_j^\intercal h_j-\delta\bu_j^\intercal\bu_j-\epsilon_j h_j^\intercal h_j,\qquad j=1,2,\dots,m
\end{equation*}
for some \(\delta_j\geq0,\epsilon_j\geq0.\)

It is straightforward to define constant passivity indices using the above definition of passivity, however, it is restrictive for switched systems to have constant indices regardless of the active subsystem. The alternative is to redefine the indices to be time-varying. As the switched system changes the active subsystem, the value of the indices change accordingly. In the case of a constant index, if the subsystems have OFP index set \(\{\rho_i\}\) (IFP index set \(\{\nu_i\}\)), then the switched system is OFP with index \(\rho=\min\{\rho_i\}\)  (IFP with index \(\nu=\min\{\nu_i\}\)). A comparison between the two cases and control design based on constant passivity indices for switched systems are presented in~\autocite{mccourt_passivity_2009}. Passivity indices are considered in~\autocite{mccourt_control_2010} for the overall switched system indices are allowed to be a function of time. In this definition, the value of the index function at any time is simply the value of the constant passivity index for the active subsystem. Formally, if each subsystem \(i\) has constant OFP index \(\rho_i\) (or IFP index \(\nu_i\)), then the overall switched system is OFP with index \(\rho(t)=\rho_\sigma\) (IFP with index \(\nu(t)=\nu_\sigma\)), where \(\sigma\) denotes the index of the active subsystem~\autocite{mccourt_control_2010}.

A notion of QSR dissipativity for discrete time switched systems that uses the multiple supply rates is provided in~\autocite{mccourt_stability_2012}. Using this notion, stability of the unforced switched system, passivity of the feedback interconnection of switched systems, and conditions for the feedback interconnection to have an stable unforced equilibrium are derived.

The results presented so far consider systems in which every mode is dissipative or passive. For more general switched systems in which some modes can be non-passive, stability has been studied in~\autocite{mccourt_control_2010,liu_stabilization_2012}. The authors in~\autocite{wang_passivity_2014} define passivity in the presence of nonpassive modes. Since the dynamics of some modes are not passive, by classical definitions, the system is not passive. However, the authors provide necessary and sufficient condition, that if the nonpassive modes are active infrequently enough, the system can still be passive. They also proved that a nonpassive system can be rendered locally passive using feedback control laws if and only if its zero dynamics are locally passive. The proposed definition in~\autocite{wang_passivity_2014} is consistent with the traditional definition of passivity in the sense that useful properties, such as stability and compositionality, are preserved.

\subsection{Passivation of Switched and Hybrid Systems} \label{sec:passivation:switched}
In~\cite{ghanbari_passivation_2016}, the M-matrix method is generalized to switched controllers consisting of \(N\) non-passive subsystems. It is assumed that a switched controller is placed in feedback configuration of a non-passive, unstable plant. The passivation matrix \(M\) is also time varying and switches between constant values whenever the controller switches. Due to the complexity of the switched system, it might be difficult to switch between different passivation matrices. 
In~\autocite{ghanbari_design_2017}, a single passivation matrix is considered. Based on~\autocite{zhao_vector_2005}, the gain of the switched controller is defined as the maximum of gains of each subsystem. Then, a single passivation matrix is designed to compensate for the lack of passivity in the system. Placing
the passivated switched controller in the negative feedback interconnection of a non-passive and unstable plant can make the closed-loop system passive and stable.

In \autocite{agarwal_feedback_2017}, a state feedback method is presented to passivate a discrete-time switched linear system to a desired strict IFP level. The importance of strict IFP, in this case, is the relation between a nonlinear switched system and its linearization. Therefore, by applying the state feedback designed with the aid of the linearized system, the nonlinear system will be passive as well.

\section{Conclusions}\label{sec:conc}
Passivity and dissipativity have shown great promise in the analysis and design of Cyber-physical systems due to their compositional properties. Passivity indices framework generalizes passivity properties to systems that may not be passive and has recently been employed in many components of CPS. In this paper, a brief outline of some of the recent advances in this area is given. In the area of networked control systems and event-triggered control, analysis and control design subject to challenges including time-varying delays, quantization, and packet dropouts were reviewed as well as trade-offs between performance and communication constraints. Design and analysis of switched and hybrid systems are also presented, especially when some subsystems are not passive,  or when the switched system is part of a networked control system. Passivation methods based on indices are also presented for different classes of systems. By necessity, only a brief outline was given here.
\printbibliography

\end{document}